\documentclass[aps,superscriptaddress,showkeys,showpacs]{revtex4}
\usepackage{amsmath,amssymb}
\usepackage{graphics,graphicx}
\usepackage{dcolumn,bm}

\newcommand{\sinp}{\affiliation{Theoretical Condensed Matter Physics Division
and Centre for Applied Mathematics and Computational Science,\\
Saha Institute of Nuclear Physics, 1/AF Bidhannagar, Kolkata 700 064, India.}}
\newcommand{\imsc}{\affiliation{Institute of Mathematical Sciences, C. I. T. Campus, Taramani, Chennai 600 113, India.}}

\begin{document}
\title{Economic Inequality: Is it Natural ?}

\author{Arnab Chatterjee}%
\email{arnab.chatterjee@saha.ac.in}
\sinp
\author{Sitabhra Sinha}%
\email{sitabhra@imsc.res.in}
\imsc
\author{Bikas K. Chakrabarti}%
\email{bikask.chakrabarti@saha.ac.in}
\sinp

\begin{abstract}
Mounting evidences are being gathered suggesting that income and wealth
distribution in various countries or societies follow a robust pattern,
close to the Gibbs distribution of energy in an ideal gas in equilibrium,
but also deviating significantly for high income groups.
Application of physics models seem to provide illuminating ideas
and understanding, complimenting the observations.
\end{abstract}



\maketitle

\noindent
We are all aware of the hard fact: neither wealth nor income 
is ever uniform for us all. Justified or not, they are unevenly 
distributed; few are rich, many are poor!
Such socioeconomic inequalities seem to be a persistent fact of 
life ever since civilization began.
Can it be that it only reflects a simple natural law, understandable from
the application of physics?

\vskip .2cm
\noindent
\textit{I. Income and wealth distributions in society}\\
Investigations over more than a century and the recent availability of
electronic databases of income and wealth distribution (ranging from
national sample survey of household assets to the income tax return data 
available from governmental agencies)
have revealed some remarkable features. Irrespective of many 
differences in culture, history, social structure, indicators of relative
prosperity (such as gross domestic product or infant mortality) and, 
to some extent, the economic
policies followed in different countries, the income distribution
seems to follow a particular universal pattern, as does the wealth
distribution: 
After an initial rise, the number density of people
rapidly decays with their income, the bulk described by a Gibbs or log-normal
distribution crossing over at the very high income range 
(for 5-10\% of the richest members of the population)
to a power law with an exponent (known as Pareto exponent) 
value between 1 and 3.
This seems to be an universal feature: from ancient Egyptian 
society~$^1$ through nineteenth century 
Europe~$^{2,3}$
to modern Japan~$^{4,5}$.
The same is true across the globe today: from the advanced capitalist
economy of USA~$^{4,5}$ to the developing economy of 
India~$^{6}$. 

The power-law tail, indicating a much higher frequency of occurrence of
very rich individuals (or households) than would be expected by extrapolating
the properties of the bulk of the distribution, was
first observed by Vilfredo Pareto~$^{2}$
in the 1890s for income distribution of several societies at very different
stages of economic development. Later, 
the wealth distribution was also seen to follow similar behavior. 
Subsequently, there have been several attempts starting around the 1950s, 
mostly by economists, to explain the genesis of the power law tail
(for a review, see Champernowne~$^{3}$).
However, most of these models involved a large number of factors that
made understanding the essential reason behind the occurrence of inequality
difficult. Following this period of activity, a relative lull followed 
in the 70s and 80s when the field lay dormant, although accurate and extensive
data were accumulated that would eventually make possible precise empirical 
determination of the distribution properties. This availability of 
large quantity of electronic data and their computational analysis has
led to a recent resurgence of interest in the problem, specifically over 
the last one and half decade.

Although Pareto~$^{2}$ and Gini~$^{7}$ 
had respectively identified the power-law tail and
the log-normal bulk of the income distribution, the demonstration of both
features in the same distribution was possibly first demonstrated by
Montroll and Shlesinger~$^{8}$ through
an analysis of fine-scale income data obtained from the US Internal Revenue
Service (IRS) for the year 1935-36. It was observed that while the top 2-3 \%
of the population (in terms of income) followed a power law with Pareto 
exponent $\nu \simeq 1.63$; 
the rest followed a log-normal distribution. Later work on
Japanese personal income data based on detailed records obtained 
from the Japanese
National Tax Administration indicated that the tail of the distribution
followed a power law with $\nu$ value that fluctuated from year
to year around the mean value of $2$~$^{9}$.
Further work~$^{10}$ showed that the power law region described the top
$10$~\% or less of the population (in terms of income), while the remaining
income distribution was well-described by the log-normal form. While the
value of $\nu$ fluctuated significantly from year to year, it was observed
that the parameter describing the log-normal bulk, the Gibrat index, remained
relatively unchanged. The change of income from year to year, i.e., the
growth rate as measured by the log ratio of the income tax paid in successive
years, was observed by Fujiwara~{\em et~al}~$^{11}$ to be also
a heavy tailed distribution, although skewed, and centered about zero.
Later work on the US income distribution based on data from IRS for the years
1997-1998, while still indicating a power-law tail (with $\nu \simeq 1.7$), 
have suggested that the the lower 95~\% of the population have income
whose distribution may be better described by an exponential
form~$^{12,13}$. The same observation has been
made for income distribution in the UK for the years 1994-1999, where the
value of $\nu$ was found to vary between 2.0 and 2.3, but the bulk
seemed to be well-described by an exponential decay.

\vskip 0.5cm
\hrule
\vskip 0.2cm
\noindent
\textbf{Box 1: \textit{Income inequality: Gini coefficient and Pareto law}}
\vskip 0.5cm

\begin{center}
\includegraphics[width=6.0cm]{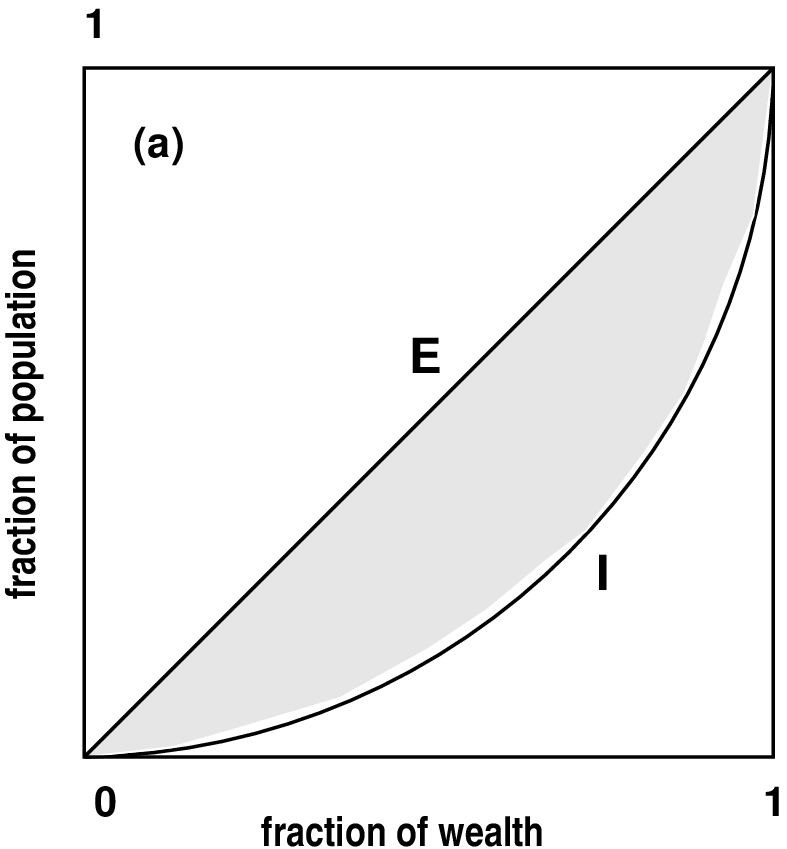}
\hskip 1cm
\includegraphics[width=6.0cm]{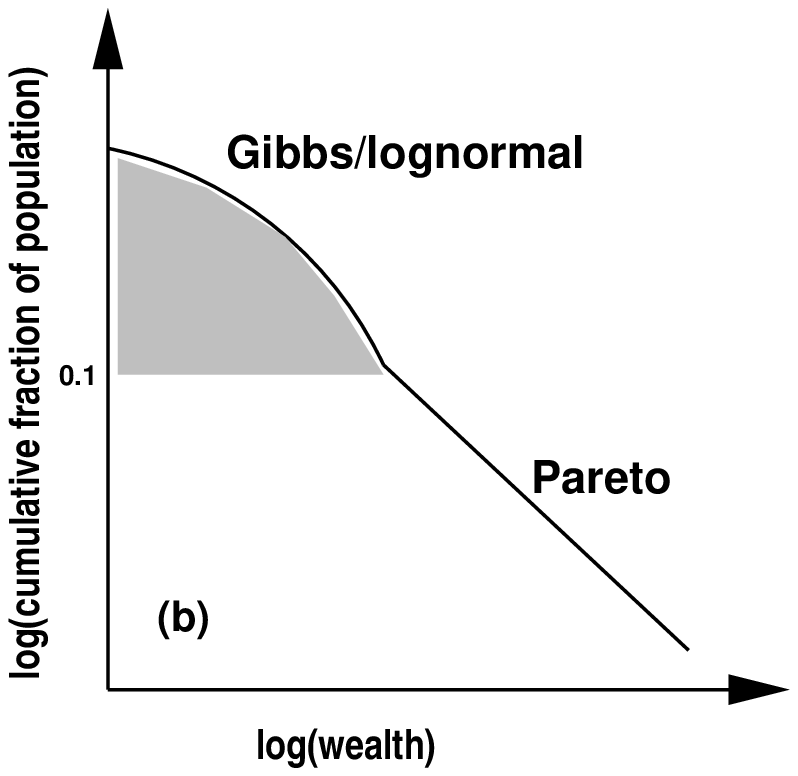}
\end{center}
\noindent
\begin{small}
(a) The Gini coefficient $G$ 
gives a measure of inequality in any income distribution and is defined as
the proportional area between the Lorenz curve ($I$, giving the
cumulative fraction of the people with the fraction of wealth) 
and the perfect equality curve ($E$, where the fraction of wealth possessed
by any fraction of population would be strictly linear):
$G=1-\frac{A_I}{A_E}$, where $A_I$ and $A_E$ are the areas
under curves $I$ and $E$ respectively. 
$G=0$ corresponds to perfect equality while $G=1$ to perfect inequality.
(b) When one plots the cumulative wealth (income) distribution against the
wealth (income), almost $90-95\%$ of the population fits the Gibbs
distribution (indicated by the shaded region in the distribution;
often fitted also to lognormal form) and for the rest (very rich) 
$5-10\%$ of the population in any country, the number density 
falls off with their wealth (income) much slowly, following a power law, 
called the Pareto law.
The second part of this law, which we do not discuss here, states
that about $40-60\%$ of the total wealth of any economy is possessed
by $5-10\%$ of the people in the Pareto tail. 
Although this seems to be qualitatively true, we do not have any 
recent data to support it.
\end{small}

\vskip 0.2cm
\hrule
\vskip 0.5cm

It is interesting to note that, when one shifts attention from the
income of individuals to the income of companies, one still observes 
the power law tail. A study of the income distribution of Japanese 
firms~$^{14}$ concluded that it follows a power law with $\nu \simeq 1$,
which is also often referred to as the Zipf's law. Similar
observation has been reported for the income distribution of US 
companies~$^{15}$.

Compared to the empirical work done on income distribution, relatively few
studies have looked at the distribution of wealth, which consist of the net
value of assets (financial holdings and/or tangible items) owned at
a given point in time.
The lack of an easily available data source for measuring wealth,
analogous to income tax returns for measuring income, means that one
has to resort to indirect methods. Levy and Solomon~$^{16}$ used
a published list of wealthiest people to generate a rank-order distribution,
from which they inferred the Pareto exponent for wealth distribution
in USA.
Refs.~$^{13}$ and~$^{17}$ used an alternative technique based on adjusted data
reported for the purpose of inheritance tax to obtain the Pareto
exponent for UK.
Another study used tangible asset (namely house area) as a measure of
wealth to obtain the wealth distribution exponent in ancient Egyptian
society during the reign of Akhenaten (14th century BC)~$^{1}$.
More recently, the wealth distribution in India at present was also
observed to follow a power law tail with the exponent varying around 
0.9~$^{6}$.
The general feature observed in the limited empirical study of
wealth distribution is that of a power law behavior for the wealthiest
5-10 $\%$ of the population, and exponential or log-normal
distribution for the rest of the population.
The Pareto exponent as measured from the wealth distribution is found
to be always lower than the exponent for the income distribution,
which is consistent with the general observation that, in market
economies, wealth is much more unequally distributed than
income~$^{18}$.
\begin{figure}
\centering\includegraphics[height=10.0cm]{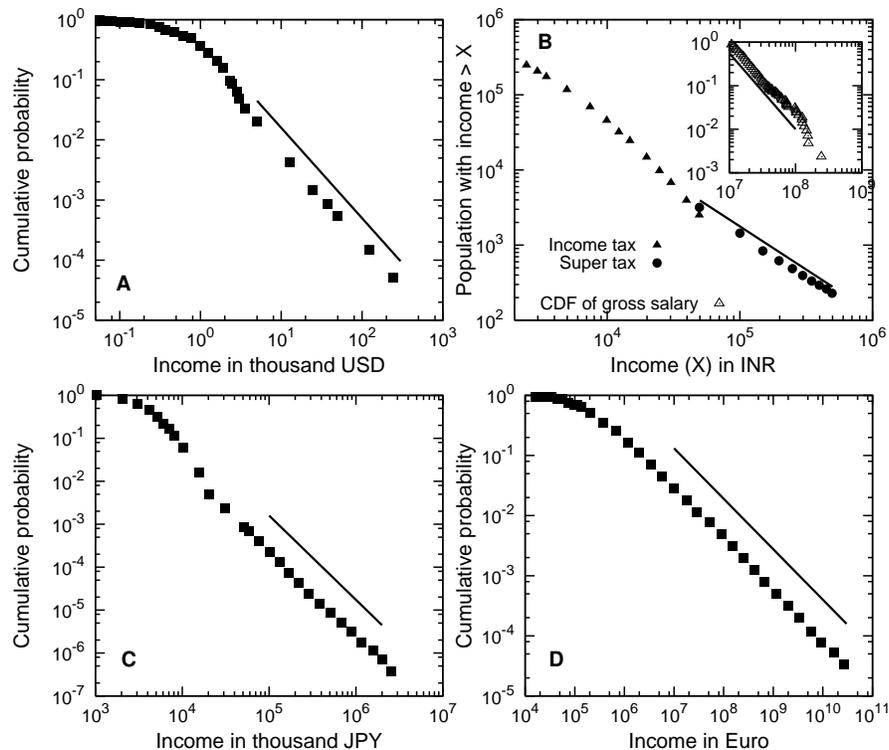}
\caption{\label{fig:realdataset}
(A) Cumulative probability ($Q(m)$) of US personal annual income ($m$)
for IRS data for 2001 (taken from Ref.~$^{19~(a)}$), 
Pareto exponent $\nu \approx 1.5$
(given by the slope of the solid line).
(B) Cumulative Income distribution in India during 1929-1930, collected from
Income Tax and Super Tax data$^{20}$.
The inset shows the cumulative distribution of the employment 
income for the top 422 salaried Indians (Business Standard survey, 2006) 
showing a power-law tail with
$\nu = 1.75 \pm 0.01$$^{21}$.
(C) Cumulative probability distribution of Japanese personal income
in the year 2000. The power law region approximately fits to 
$\nu=1.96$ (data from Ref.~$^{19~(b)}$). 
(D) Cumulative probability distribution of
firm size (total-assets) in France in the year 2001 for 669620 firms.
The power law region approximately fits to $\nu=0.84$
(data from Ref.~$^{19~(b)}$).
}
\end{figure}
The striking regularities (see Fig.~\ref{fig:realdataset}) 
observed in the income distribution for different countries, have 
led to several new attempts at 
explaining them on theoretical grounds. Much of the current impetus is
from physicists' modelling of economic behavior in analogy with
large systems of interacting particles, as treated, e.g., in the kinetic
theory of gases. According to physicists working on this problem, the
regular patterns observed in the income (and wealth) distribution may
be indicative of a natural law for the statistical properties of a
many-body dynamical system representing the entire set of economic 
interactions in a society, analogous to those previously derived for
gases and liquids. By viewing the economy as a thermodynamic system,
one can identify the income distribution with the distribution of
energy among the particles in a gas.
In particular, a class of kinetic exchange models have provided a simple
mechanism for understanding the unequal accumulation of assets.
Many of these models, while simple from the perspective of economics,
has the benefit of coming to grips with the key factor in socioeconomic
interactions that results in very different societies converging to
similar forms of unequal distribution of resources (see 
Refs.~$^{4,5}$, which consists of a collection of 
large number of technical papers in this field; see 
also~$^{22-24}$ 
for some popular discussions and also criticisms).

\vskip .2cm
\noindent
\textit{II. A simple ideal gas like model}\\
Think of an exchange game like the following in an economy where the
different commodities are not being explicitly considered, but rather
their value in terms of an uniform asset (money). In such an {\em asset
exchange} game, there are $N$ players participating, with each player
having an initial capital of one unit of money. $N$ is very large,
and total money $M=N$ remains fixed over the game as does the number of
players $N$.

(a) In the simplest version, the only allowed move at any time is that two
of these players are randomly chosen and they decide to divide their pooled
resources randomly among them. As no debt is allowed, none of the players
can end up with a negative amount of assets. 
As one can easily guess, the initial delta function
distribution of money (with every player having the same amount) gets 
destabilized with such moves and the state of perfect equality, where every
player has the same amount, disappears quickly.
Let us ask, what will be the eventual
steady state distribution of assets among the players after many such moves?
The answer is well established in physics for more than a century --- soon,
there will be a stable asset distribution and it will be the 
Gibbs distribution:
$P(m) \sim \exp[-m/T]$, where the parameter $T=M/N$ corresponds to the
average money owned by an agent~$^{25-27}$.

(b) Now think of a modified move in this game: each player `saves' a fraction
$\lambda$ of his/her total assets during every step of the game, the rest 
being pooled and randomly divided with the other (randomly chosen) player. 
If everybody saves the same fraction $\lambda$, what is the steady state
distribution of assets after a large number of such moves?
It becomes Gamma-function like, whose 
parameters of course depend on $\lambda$:
$P(m) \simeq m^\alpha \exp [-m/T(\lambda)]$; 
$\alpha=3 \lambda/(1-\lambda)$, 
(see~$^{27,28}$).
Angle, utilizing a different stochastic model, arrived at somewhat
similar (numerical) results, considerably 
earlier~$^{29,30}$.
Although qualitative explanation and limiting results for $\lambda \to 0$
or $\lambda \to 1$ are easy to obtain,
no exact treatment of this problem is available so far.

\vskip 0.5cm
\hrule
\vskip 0.2cm
\noindent
\textbf{Box 2: \textit{Kinetic theory of ideal gas: Gibbs and 
Maxwell-Boltzamann distributions}}
\vskip 0.5cm
\begin{center}
\includegraphics[width=6.0cm]{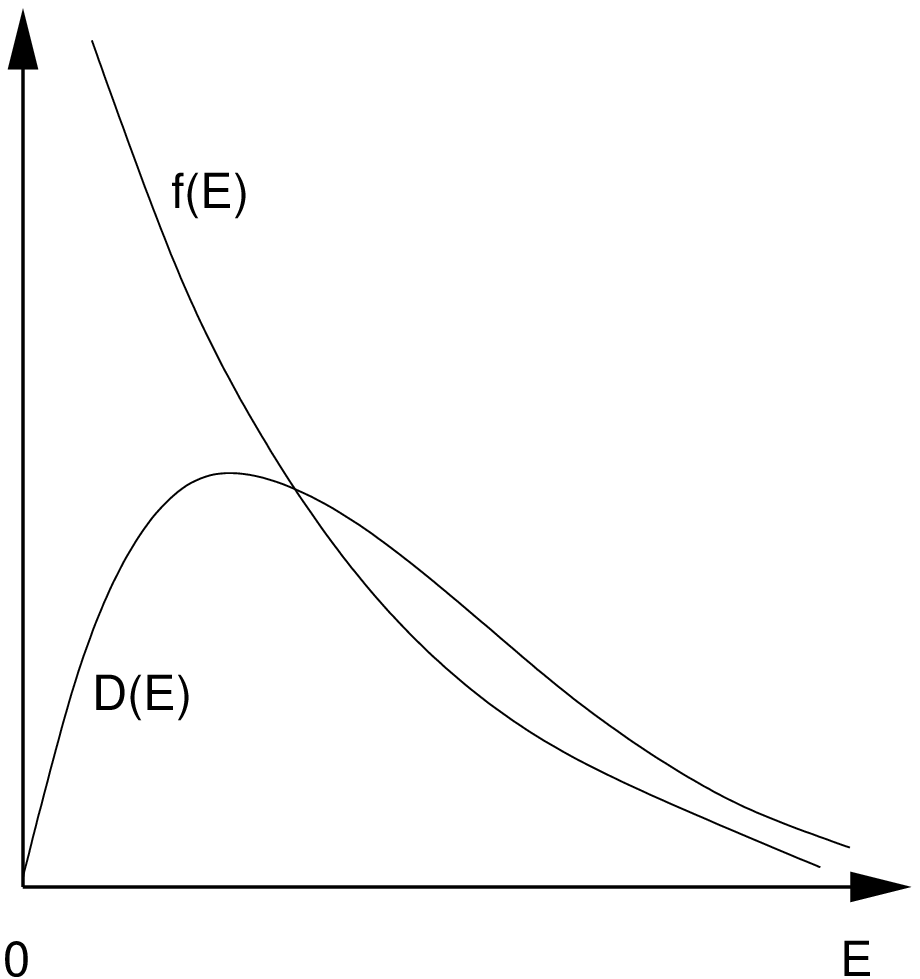}
\end{center}

\noindent
\begin{small}
In a classical ideal gas in thermodynamic equilibrium, the state variables
like pressure ($P$), volume ($V$) and the absolute temperature ($T$) 
maintains a very simple relationship $PV=NkT$. Here $N$ is the number
of basic constituents (atoms or molecules; $N \sim$ Avogadro numer 
$\sim 10^{23}$) and $k$ is a constant called Boltzmann constant.
Statistical mechanics of ideal gas, also called the kinetic theory
of gas, intends to explain the above gas law in terms of the constituents'
mechanics or kinetics. According to this picture, for a classical ideal
gas, each constituent is a Newtonian particle and they undergo random
elastic collisions (which conserve kinetic energy $E$) among themselves
and the walls of the container. These collisions eventually set up
a non-uniform (kinetic) energy distribution $D(E)$ among the constituents,
called the Maxwell-Boltzmann distribution:
$D(E)=f(E)g(E)$, where $g(E)$ ($\sim \sqrt{E}$ here for an ideal gas in
a 3-dimensional container) is called the density of states and comes from
mechanics (of free or noninteracting particles of the ideal gas), 
and $f(E)$ ($\sim \exp(-E/kT)$) is called the Gibbs distribution
and comes from the statistical mechanics (result of averages over random
scattering events).
Identifying the pressure $P$ as the average (over the distribution
$D(E)$) rate of change of momentum of the gas particles on unit area
of the container (where the energy $E$ is proportional to the square
of the momentum), and the temperature $T$ as the average (over the 
distribution $D(E)$) energy, one immediately gets the above mentioned
gas law (relating $P$, $V$ and $T$).
\end{small}
\vskip 0.2cm
\hrule
\vskip 0.5cm

\begin{figure}
\centering\includegraphics[height=13.0cm]{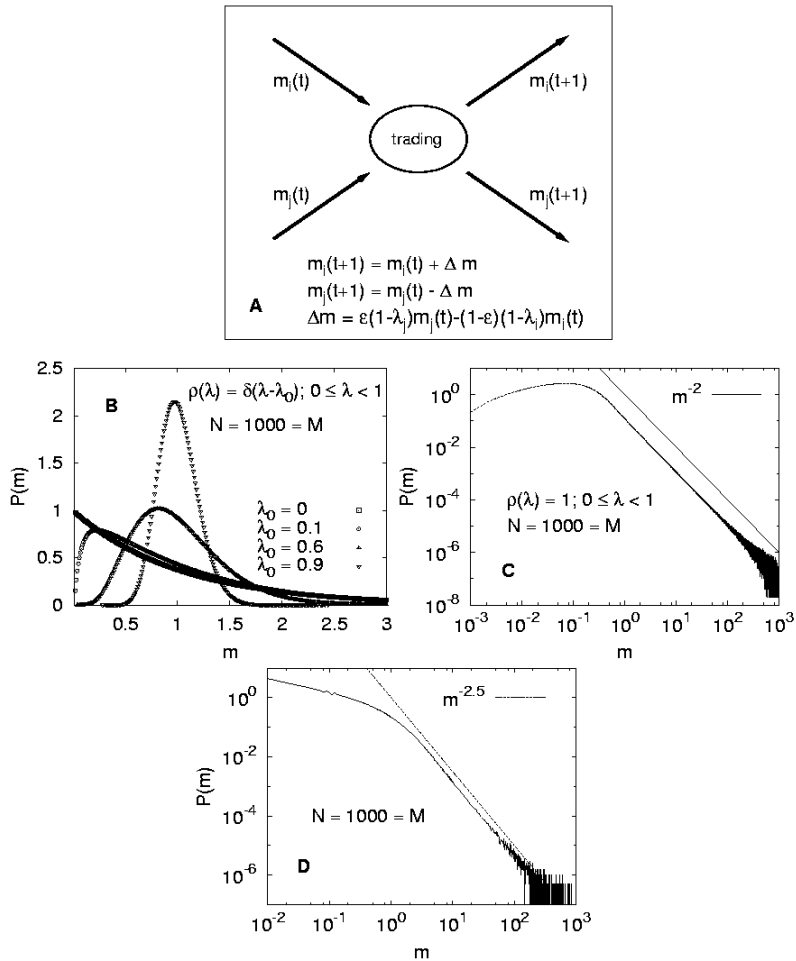}
\caption{\label{fig:simul}
(A) The trading markets can be easily modelled to be composed 
of two-body scatterings as shown above:
The money $m_i(t)$ of an agent $i$ at time $t$ 
changes due to trading/scattering
with a random agent $j$ in the market; the scattering locally conserves
the total money. Each agent saves a fraction $\lambda_i$ of its money $m_i(t)$
at that time $t$ and the same is true for the other, and the rest of the money
$(1-\lambda_i)m_i(t) + (1-\lambda_j)m_j(t)$ is shared randomly
($\epsilon$ is a random fraction between $0$ and $1$).
We assume $\epsilon$ to be an annealed variable (changes with trading 
or time), while $\lambda$ are quenched variables (do not change with time).
$\lambda_i$ can of course change from agent to agent, given by its
distribution $\rho(\lambda)$. (B) For uniform $\lambda$, a Gamma distribution
$P(m)$ for money occurs, (C) while for a white distribution of $\lambda$,
a Pareto law $P(m) \sim m^{-2}$ (i.e. Pareto exponent $\nu=1$) sets in.
Asset distribution in the
asymmetric asset exchange game where the players have different thrift
values (randomly chosen from an uniform distribution over the unit interval)
also exhibits a power law tail, as shown in (D), with Pareto
exponent $\nu \simeq 1.5$.
In comparing with the cumulative probability $Q(m)$ in 
Fig.~\ref{fig:realdataset}, one should note that $Q(m)$ is given by 
$\int_m^\infty P(m) dm$.}
\end{figure}

(c) What happens to the steady-state asset distribution among these players if
$\lambda$ is not the same for all players but is different for different
players? Let the distribution $\rho(\lambda)$ of saving propensity
$\lambda$ among the agents be such
that $\rho(\lambda)$ is non-vanishing when $\lambda \to 1$.
The actual asset distribution $P(m)$ in such a model will depend on the 
saving propensity distribution
$\rho(\lambda)$, but for all of them the asymptotic form of the
distribution will become
Pareto-like: $P(m) \sim m^{-(1+\nu)}$; $\nu=1$ for 
$m \to \infty$~$^{31-33}$.
This is valid for all such distributions (unless $\rho(\lambda) \propto
(1-\lambda)^\delta$, when 
$P(m) \sim m^{-(2+\delta)}$~$^{32}$).
However, for variation of $\rho(\lambda)$ such that $\rho(\lambda) = 0$
for $\lambda < \lambda_0$ and $\rho(\lambda) \ne 0$ for
$\lambda_0 < \lambda < 1$, one will get an initial Gamma function form
for $P(m)$ for small and intermediate values of $m$, with parameters
determined by $\lambda_0$ ($\ne 0$), and this distribution will
eventually become Pareto-like for
$m \to \infty$ with $\nu =1$ (see Fig.~\ref{fig:simul}; 
cf.~$^{31-33}$).
Analytical understanding is now available~$^{34,35}$ and Ref.~$^{36}$ 
gives a somewhat rigorous analytical treatment of this problem.

It may be mentioned that there are a large number of random multiplicative
asset exchange models (see. e.g,~$^{37,38}$) to explain the Pareto
(power-law) tail of the wealth or income distribution. The advantage of the 
kind of model discussed above is that it can accommodate all the essential 
features of $P(m)$ for the entire range of $m$, not only the Pareto tail.

(d) One can of course argue that the random division of pooled assets
among players is not a realistic approximation of actual trading carried
out in society. As Hayes~$^{22}$ points out, in most exchanges
between an individual and a large company, it is unlikely that the
individual will end up with a significant fraction of the latter's assets.
A strict enforcement of this condition leads to a new type of game, the
{\em minimum exchange} model, where the maximum amount that can change
hands over a move, is a fraction of the poorer player's assets. Although
the change in the rules does not seem significant from the simple
random exchange game, the outcome is astonishingly different: in the steady
state, one player ends up with all the assets. 

If we now relax the condition that the richer player does not completely
dictate the terms of exchange, so that the amount exchanged need not be
limited by the total asset owned by the poorer player, we arrive at a game
which is asymmetric in the sense of generally favoring the player who is 
richer than the other, but not so much that the richer player dominates
totally. Just like the previously defined savings propensity for a player,
one can now define `thrift' $\tau$, which measures the ability of a player 
to exploit
its advantage over a poorer player~$^{20}$. For the two extreme
cases of minimum ($\tau = 0$) and maximum ($\tau = 1$) thrift, one gets back 
the random asset exchange
and minimum asset exchange models, respectively. However, close to 
the maximum limit, at the transition between the two very different
steady-state distributions given by the two models, we see a power-law 
distribution!
As in the case of $\lambda$, we can now consider the case when instead
of having the same $\tau$, different players are endowed with different
thrift abilities. For such heterogeneous thrift assignment in the population,
where $\tau$ for each player is chosen from a random distribution, the
steady-state distribution reproduces the entire 
range of observed distributions
of income (as well as wealth) in the society: the tail follows a power law, 
while the bulk is described by an exponential distribution. 
The tail exponent depends on the distribution of $\tau$, with the value
of $\nu = 1.5$ suggested originally by Pareto, obtained for the simplest 
case of uniform distribution of $\tau$ between [0,1]
(see Fig.~\ref{fig:simul}~D). However, even extremely
different distributions of $\tau$ (e.g., U-shaped) always produces a power-law
tailed distribution that is exponentially decaying in the bulk, underlining
the robustness of the model in explaining inequality.

\vskip .2cm
\noindent
\textit{III. An extension}\\
A major limitation of these asset exchange models considered earlier
(and summarized above in Sec.~\textit{II}) is that
it does not make any explicit reference to the commodities exchanged whose
asset values we were considering so far and to the constraints they impose.
We have also studied~$^{39}$ the effect of explicitly introducing
a single non-consumable commodity (which is bought and sold in terms of
money) on the asset distributions in the steady state. 
Here, again two of the agents are arbitrarily chosen for interaction
(or trading) and the commodity exchanged for money, provided of course
the two agents have the required amounts of commodity and money (since
no credit purchases are allowed).
Otherwise, no exchange take place
and a new pair of agents is chosen. The global price of the commodity
(ratio of total money to total amount of the commodity in the market) is 
normalized but
have temporal fluctuations.
Here, we distinguish between money and wealth; wealth of any agent is
composed of money and the money equivalent of the commodity with the agent.
In spite of many significant effects,
the general feature of Gamma-like form of the asset
distributions (for uniform $\lambda$) and the power law tails
(for random $\lambda$) for both money and wealth, with identical exponents, 
are seen to remain unchanged. 

These studies indicate that the precise studies (theories) for the asset
exchange models are extremely useful and relevant.
Also this helps us to address the question of identifying a money-like 
asset with
wealth in simple asset exchange models and suggests that the absurd
simplicity can be relaxed, yet the quantitative features are not affected.

\vskip .2cm
\noindent
\textit{IV. Relevance of gas like models}

\noindent
All these gas-like models of trading markets are based on the assumption of
(a) asset conservation (globally in the market; as well as locally
in any trading)
and
(b) stochasticity.
Questions on the validity of these points are very natural and have been 
raised~$^{4,5,40}$. We now forward some arguments in 
their favor.

\vskip .2cm
\noindent
\textit{(a) Asset conservation:}
If we view the trading as scattering processes, one can see the
equivalence. Of course, in any such `asset exchange' trading process,
one receives some profit or service from the other and this does
not appear to be completely random, as assumed in the models.
However, if we concentrate only on the `cash' exchanged
(even using Bank cards!), every trading is an asset conserving one
(like the elastic scattering process in physics!)
As discussed in Sec.~\textit{III}, conservation of asset can be 
extended to that of total wealth (including money) and relaxed,
as given by the temporally fluctuating price
(effectively allows for slight relaxation over this conservation), yet
keeping the overall distribution same
(with unchanged $\nu$ value)~$^{39}$.
It is also important to note that the frequency of asset exchange
in such models define a time scale in which the total asset
in the market does not change. In real economies, the total asset
changes much slowly, so that in the time scale of exchanges,
it is quite reasonable to assume the total asset to be conserved in
these exchange models.

\vskip .2cm
\noindent
\textit{(b) Stochasticity:}
But, are these trading random? Surely not, when looked upon from
individual's point of view. When one maximizes his/her utility by
money exchange for the $p$-th commodity, he/she may choose to
go to the $q$-th agent and for the $r$-th commodity he/she will go to the
$s$-th agent. But since $p \ne q \ne r \ne s$ in general, when viewed
from a global level, these trading/scattering events will all look
random (although for individuals this is a defined choice or utility
maximization).
It may be noted in this context that in the stochastically formulated
ideal gas models in physics (developed in late 1800/early 1900),
physicists already knew for more than a hundred years that
each of the constituent particle (molecule) follows a precise equation
of motion, namely that due to Newton.
The assumption of stochasticity in asset exchange models, even though 
each agent
might follow an utility maximizing strategy (like Newton's equation
of motion for molecules), is therefore not very unusual in the context.

\vskip .2cm
\noindent
\textit{(c) Support from economic data:}
Analysis of high quality income data~$^{41}$ from UK and USA
show peaked Gamma distributions for the low and middle income ranges, 
which suggest a strong case in favor of the models discussed in 
\textit{II(b)-(d)}~$^{20,27,32}$.
This has already been seen in studies of isolated groups of similar 
individuals, and has been modelled in similar fashion~$^{29,30}$.

\vskip .2cm
\noindent
\textit{V. Concluding remarks}

\noindent
The enormous amount of data available on the income and wealth distribution
of various countries clearly establish a robust feature:
Gamma (or log-normal) distribution for the majority (almost 90-95\%),
followed by a Pareto power law (for the richest 5-10\% of the population),
as seen in Fig.~\ref{fig:realdataset}.
We show that this `natural' behavior of income inequality comes from
a simple `scattering picture' of the market (see Fig.~\ref{fig:simul}~A),
when the agent in the market have got random saving propensity.
Models studied in physics (in kinetic theory of gases), 
more than a hundred years ago, help us in formulating and understanding
these `natural' behavior of the markets.

\vskip .5cm
\noindent
\textbf{Acknowledgments}

\noindent
The authors are grateful to 
P.~Bhattacharyya, A.~Chakraborti, S.~S.~Manna, S.~Marjit, S.~Pradhan and
R.~B.~Stinchcombe for collaborations at various stages of the study.
Useful discussions with
J.~Angle, A.~S.~Chakrabarti, A.~Das, M.~Gallegati, A.~Kar~Gupta, 
T.~Lux, P.~K.~Mohanty, M.~Marsili, R.~K.~Pan, S.~Raghavendra, 
P.~Richmond, D.~Stauffer, S.~Subramaniam, V.~M.~Yakovenko and S.~Yarlagadda 
are also acknowledged.


\begin{thebibliography}{99}

\bibitem{Egypt}
Abul-Magd, A. Y., \textit{Phys. Rev. E}, 2002, \textbf{66}, 057104.

\bibitem{cc:Pareto:1897}
Pareto, V., \textit{Cours d'economie Politique},
F. Rouge, Lausanne and Paris, 1897.

\bibitem{cc:Champernowne:1999}
Champernowne, D. G., \textit{The Economic Journal}, 1953, \textbf{63}, 318;
Champernowne, D. G., Cowell, F. A., \textit{Economic Inequality and Income
Distribution}, Cambridge University Press, Cambridge, 1999.

\bibitem{cc:EWD05}
\textit{Econophysics of Wealth Distributions},
(eds. Chatterjee, A., Yarlagadda, S., Chakrabarti, B. K.),
Springer Verlag, Milan, 2005.

\bibitem{Wileybook}
KarGupta, A., pp.~161-190 and Richmond, P., Hutzler, S., Coelho, R.,
Repetowicz, P., pp.~131-159,
in \textit{Econophysics and Sociophysics: Trends and Perspectives},
(eds. Chakrabarti, B. K., Chakraborti, A., Chatterjee, A.), 
Wiley-VCH, Berlin, 2006.

\bibitem{cc:Sinha:2006}
Sinha, S., \textit{Physica A}, 2006, \textbf{359}, 555.

\bibitem{Gini}
Gini, C., \textit{The Economic Journal}, 1921, \textbf{31}, 124.

\bibitem{MontrollShlesinger}
Montroll, E. W., Shlesinger, M. F., \textit{Proc. Natl. Acad. Sci. USA}, 
1982, \textbf{79}, 3380.

\bibitem{Aoyamaetal}
Aoyama, H., Souma, W., Nagahara, Y., Okazaki, M. P., Takayasu, H., 
Takayasu, M., \textit{Fractals}, 2000, \textbf{8}, 293.

\bibitem{Souma}
Souma, W., \textit{Fractals}, 2000, \textbf{9}, 463.

\bibitem{Fujiwaraetal}
Fujiwara, Y., Souma, W., Aoyama, H., Kaizoji, T., Aoki, M.,
\textit{Physica A}, 2003, \textbf{321}, 598.

\bibitem{DraguYakov01a}
Dr\u{a}gulescu, A. A., Yakovenko, V. M., 
\textit{Eur. Phys. J. B}, 2001, \textbf{20}, 585.

\bibitem{DraguYakov01b}
Dr\u{a}gulescu, A. A., Yakovenko, V. M.,
\textit{Physica A}, 2001, \textbf{299}, 213.

\bibitem{Okuyamaetal}
Okuyama, K., Takayasu, M., Takayasu, H.,
\textit{Physica A}, 1999, \textbf{269}, 125.

\bibitem{Axtell}
Axtell, R. L.,
\textit{Science}, 2001, \textbf{293}, 1818.

\bibitem{Lev97}
Levy, M., Solomon, S.,
\textit{Physica A}, 1997, \textbf{242}, 90.

\bibitem{Coe04}
Coelho, R., N\'{e}da, Z., Ramasco, J. J., Santos, M. A.,
\textit{Physica A}, 2005, \textbf{353}, 515.

\bibitem{Sam01}
Samuelson, P. A., Nordhaus, W. D.,
\textit{Economics}, 17th ed, McGraw Hill, New York, 2001.

\bibitem{YakoFujidata}
(a) Silva, A. C., Yakovenko, V. M. in Ref.~\cite{cc:EWD05};
(b) Fujiwara, Y. in Ref.~\cite{cc:EWD05}.

\bibitem{cc:sinha05}
Sinha, S., in Ref.~\cite{cc:EWD05}, p.~177-183.

\bibitem{aboutData}
In spite of the best of our efforts in collecting the equivalent data from
the Income Tax Department of the Government of India or the Reserve Bank 
of India, we are unable to give or compare with any better data. 

\bibitem{cc:Hayes}
Hayes, B.,
\textit{Am. Scientist}, 2002, \textbf{90}, 400.

\bibitem{cc:Hogan}
Hogan, J., 
\textit{New Scientist}, 2005, 12 March, 6.

\bibitem{cc:Ball}
Ball, P., 
\textit{Nature}, 2006, \textbf{441}, 686; Editorial, \textit{Nature}, 2006,
\textbf{441}, 667.

\bibitem{cc:marjit}
Chakrabarti, B. K., Marjit, S.,
\textit{Indian J. Phys. B}, 1995, \textbf{69}, 681.

\bibitem{Dragulescu:2000}
Dr\u{a}gulescu, A. A., Yakovenko, V. M., 
\textit{Eur. Phys. J. B}, 2000, \textbf{17}, 723.

\bibitem{cc:Chakraborti:2000}
Chakraborti, A., Chakrabarti, B. K.,
\textit{Eur. Phys. J. B}, 2000, \textbf{17}, 167.

\bibitem{cc:Patriarca:2004}
Patriarca, M., Chakraborti, A., Kaski, K.,
\textit{Phys. Rev. E}, 2004, \textbf{70}, 016104.

\bibitem{cc:Angle:1986}
Angle, J.,
\textit{Social Forces}, 1986, \textbf{65}, 293.

\bibitem{cc:Angle:2006}
Angle, J.,
\textit{Physica A}, 2006, \textbf{367}, 388.

\bibitem{cc:Chatterjee:2003}
Chatterjee, A., Chakrabarti, B. K., Manna, S. S.,
\textit{Phys. Scr. T}, 2003, \textbf{106}, 36.

\bibitem{cc:Chatterjee:2004}
Chatterjee, A., Chakrabarti, B. K., Manna, S. S.,
\textit{Physica A}, 2004, \textbf{335}, 155.

\bibitem{cc:Chakrabarti:2004}
Chakrabarti, B. K., Chatterjee, A.,
in \textit{Application of Econophysics},
(ed. Takayasu, H.) Springer, Tokyo, 2004, pp. 280-285.

\bibitem{cc:Repetowicz:2005}
Repetowicz, P., Hutzler, S., Richmond, P.,
\textit{Physica A}, 2005, \textbf{356}, 641.

\bibitem{cc:Chatterjee:2005}
Chatterjee, A., Chakrabarti, B. K., Stinchcombe, R. B.,
\textit{Phys. Rev. E}, 2005, \textbf{72}, 026126.

\bibitem{cc:Mohanty:2006}
Mohanty, P. K.,
\textit{Phys. Rev. E}, 2006, \textbf{74}, 011117.

\bibitem{cc:levysolomon}
Levy, M., Solomon, S.,
\textit{Int. J. Mod. Phys. C}, 1995, \textbf{7}, 595.

\bibitem{cc:othermodels}
Sinha, S.,
\textit{Phys. Scr. T}, 2003, \textbf{106}, 59.

\bibitem{cc:commo}
Chatterjee, A., Chakrabarti, B. K.,
\textit{Eur. Phys. J. B}, 2006, \textbf{54}, 399.

\bibitem{cc:Gallegati:2006}
Gallegati, M., Keen, S., Lux, T., Ormerod, P.,
\textit{Physica A}, 2006, \textbf{370}, 1.

\bibitem{WillisMimkes}
Willis, G. M., Mimkes, J., 2004, arXiv:cond-mat/0406694.


\end{thebibliography}
\end{document}